\documentclass[aps,prb,twocolumn,floatfix,showpacs]{revtex4}

\usepackage{amsfonts, amsmath, amssymb,graphicx}
\usepackage[colorlinks,citecolor=blue,linkcolor=blue,urlcolor=blue]{hyperref}

\begin{document}

\title{Topological pumping in the one-dimensional Bose-Hubbard model} 

\author{Davide Rossini}
\affiliation{NEST, Scuola Normale Superiore and Istituto Nanoscienze-CNR, I-56126 Pisa, Italy}
\author{Marco Gibertini}
\affiliation{NEST, Scuola Normale Superiore and Istituto Nanoscienze-CNR, I-56126 Pisa, Italy}
\author{Vittorio Giovannetti}
\affiliation{NEST, Scuola Normale Superiore and Istituto Nanoscienze-CNR, I-56126 Pisa, Italy}
\author{Rosario Fazio}
\affiliation{NEST, Scuola Normale Superiore and Istituto Nanoscienze-CNR, I-56126 Pisa, Italy}

\begin{abstract}
  By means of time-dependent density matrix renormalization group calculations we study topological 
  quantum pumping in a strongly interacting system. The system under consideration is described by the 
  Hamiltonian of a one-dimensional extended  Bose-Hubbard model in the presence of a correlated hopping 
  which breaks lattice inversion symmetry. This model has been predicted to support topological pumping 
  [E. Berg, M. Levin, and E. Altman, Phys. Rev. Lett. {\bf 106}, 110405 (2011)]. The pumped charge is 
  quantized and of topological nature.  We provide a detailed analysis of the finite-size-scaling 
  behavior of the pumped charge and its deviations from the quantized value.  Furthermore we also 
  analyze the non-adiabatic corrections due to the finite frequency of the modulation. 
  We consider two configurations: a closed ring where the time-dependence of the parameter 
  induces a circulating current, and a finite open-ended chain where particles are dragged from one edge
  to the opposite edge, due to the pumping mechanism induced by the bulk.
\end{abstract}



\maketitle

\section{Introduction}   \label{Introduction}

A charge/spin flow can take place, in the absence of any direct bias, by a periodical modulation 
of some parameters (magnetic fields, coupling constants, external gates,~$\dots$) affecting 
the dynamics of the system~\cite{Tho83}. This transport mechanism, known as quantum pumping, 
is intimately related to the properties of the instantaneous eigenvalues and eigenfunctions 
of the system Hamiltonian. Pumping is adiabatic if the variation of the parameters is slow 
as compared to the characteristic time scales of the system. 
Since the pioneering work of Thouless in 1983~\cite{Tho83} pumping has been attracting an intense 
theoretical interest~\cite{brouwer98,pumpingth} and experimental interest~\cite{pumpingexp}. 
In the scattering approach to quantum transport~\cite{brouwer98}, the pumped charge in an adiabatic 
cycle can be expressed in terms of an integral in parameter space which does not depend 
on the evolution period $T$. 
Accordingly, adiabatic quantum pumping appears to possess an intrinsic geometric character 
which allows us to relate it with the Berry phase accumulated by the system wave function 
during the cyclic evolution~\cite{geometric}. Interestingly enough, under proper conditions, 
the quantum pumping can acquire also a topological nature which yields to a quantization 
of the pumped charge/spin per cycle. 
This effect is connected to the appearance, in the Hamiltonian spectrum, of gapless points 
enclosed by the adiabatic loop in parameter space~\cite{Tho83,Sim83}.
Conditions for the quantization of the pumped charge have been discussed in Ref.~\onlinecite{makhlin}, 
while, for the case of spin transport, a classification of topological non-interacting pumps 
has been given in Ref.~\onlinecite{meidan}.

The study of quantum pumping has been extended also to interacting systems~\cite{Niu84,Avr85,interacting,meidan2}.
Here the Hamiltonian gapless points responsible for topological pumping stem from many-body effects 
and are typically associated to quantum phase transitions. 
This suggests that quantized adiabatic transport can be achieved by evolving the system 
around a loop in parameter space enclosing a critical point.
Recently it has been suggested, with the support of an analysis based on bosonization, that quantized 
pumping occurs in systems of interacting lattice bosons at integer filling in one dimension~\cite{Ber11}. 
When both on-site and nearest-neighbor interactions are taken into account, a critical point 
arises separating two distinct insulating phases~\cite{Tor06}. Encircling this point in parameter 
space adiabatically is expected to pump exactly one boson per cycle. 

The interesting result given in Ref.~\onlinecite{Ber11} calls for a detailed analysis 
of the pumping mechanism in the extended Bose-Hubbard model. Here we tackle this problem by means 
of numerical time-dependent density matrix renormalization group (t-DMRG) simulations~\cite{Sch11}. 
We compute the number of pumped particles per cycle as a function of the system size and show that 
the scaling is consistent with unit quantized pumping when the adiabatic loop encloses 
the critical point, while it otherwise vanishes. 
We show that the evolution period $T$ does not affect the pumping properties of the system, 
provided it is chosen so that the system remains adiabatic with respect to the minimum instantaneous gap. 
The topological character of the pumping mechanism is tested by considering different loops 
and assessing the independence of the number of pumped particles on the specific trajectory 
in parameter space. A quantitative analysis of finite-size and non-adiabatic effects is a necessary 
and important step towards a possible experimental observation of topological pumps.  

Pumping in the Bose-Hubbard model can be analyzed in two different configurations. 
One can consider a ring geometry. Here the modulation of parameters will result in a persistent 
pumped current. 
Alternatively one can place the interacting region between two (non-interacting) reservoirs, 
in which case particles are pumped from one lead to the other. 
We will consider in great details the first case and comment on the second setup at the end of the paper.
Concerning the latter case, we will provide results about the simulation of a pure interacting system
in an open-ended geometry, showing that the ultimate effect of the pumping mechanism after one loop
is to drag particles from one edge to the opposite edge of the chain, leaving the bulk unaffected.

The paper is organized as follows. In the next section we define the model and summarize 
the salient features of its phase diagram which are of relevance for the topological pumping. 
Section~\ref{sec:method} is devoted to the description of the method and of all the important
technical details we employed to study both static and dynamical properties of the model.
We then discuss quantitatively the behavior of the gaps at the critical point and along the loops in
parameter space we are going to consider  (Sec.~\ref{sec:gaps}).
In Sec.~\ref{sec:pump} we explain how we compute the pumped charge. 
We first consider the ring setup (Sec.~\ref{sec:ring}), discussing the deviations 
from perfect quantization due to the finite number of sites and the finite modulation period. 
Then we address the case of a transport experiment in an open-ended interacting system, 
where particles are pumped from one end of the chain to the other (Sec.~\ref{sec:obc}). 
The last section is devoted to the conclusions of this work.

\section{The model}   \label{sec:model}

The model we consider consists of a collection of interacting lattice bosons on a chain
of $L$ sites at unit filling ({\it i.e.}, with the total number of particles set to be constant and equal 
to the number of sites).
For this system we assume an Hamiltonian given by the sum of two contributions:
\begin{equation}\label{eq:fullH}
  \hat{\cal H} = \hat{\cal H}_{\rm 0} + \hat{\cal H}_{\lambda}~. 
\end{equation}
The first term describes an extended Bose-Hubbard model, defined by the expression
\begin{eqnarray}\label{eq:EBH}
  \hat{\cal H}_{\rm 0} = \sum_{j=1}^{L} \left[ -(\hat{b}^{\dag}_{j} \hat{b}_{j+1}  \hspace{-0.07cm} 
    + \hspace{-0.07cm} {\rm h.c.}) + \frac{U}{2}  \hat{n}_j (\hat{n}_j \hspace{-0.07cm}-\hspace{-0.07cm}1)
    + V  \hat{n}_j \hat{n}_{j+1} \right] \hspace{-1.mm},
\end{eqnarray}
where, for $j = 1, \cdots, L$, the operator $\hat{b}^{\dag}_{j}$ ($\hat{b}_{j}$) creates (destroys) a boson 
at lattice site $j$ ($\hat{n}_{j} = \hat{b}^{\dag}_j \hat{b}_j$ being the associated occupation operator), 
while, depending on the chosen boundary conditions, $\hat{b}_{L+1}$ is set equal 
to $\hat{b}_{1}$ (ring geometry) or to $0$ (open chain geometry).
In Eq.~(\ref{eq:EBH}) $U$ and $V$ respectively denote the on-site and nearest-neighbor interaction 
strengths in units of the hopping strength, which hereafter implicitly sets the energy scale. 

\begin{figure}[!t]
  \includegraphics[width=0.9\linewidth]{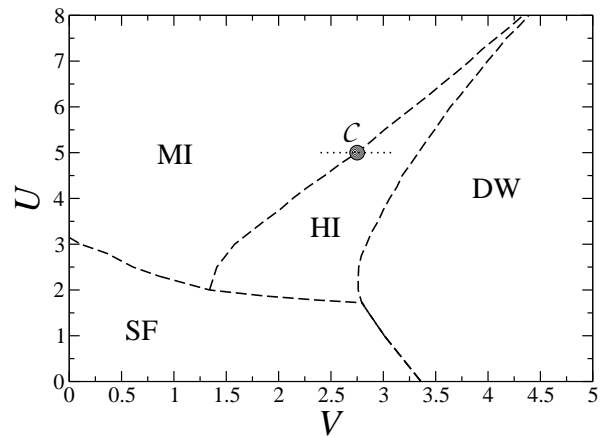} 
  \caption{Phase diagram of the one-dimensional extended Bose-Hubbard model defined by 
    the Hamiltonian $\hat{\cal H}_{\rm 0}$, at unit filling~\cite{trunc}.
    The various regions separated by lines denote the following phases: 
    MI = Mott insulator, HI = Haldane insulator, SF = superfluid, DW = density wave. 
    The boundaries between the three insulating phases have been obtained by computing
    the string ${\cal O}_{\rm string}$ and the staggering order parameter ${\cal O}_{\rm DW}$, 
    while the SF phase has been located by analyzing the closure of the ground-state charge gap 
    (see Ref.~\onlinecite{Ros12} for details). 
    In the whole paper we concentrate on the region at $U = 5$ and close to the critical point 
    $\mathcal {C}$ at $V_{\rm c}\approx 2.7$ (filled circle), as indicated by an horizontal dotted line. 
    The values of $U$ and $V$ are expressed in units of the hopping strength.}
  \label{fig:diagram} 
\end{figure}

The zero-temperature phase diagram of $\hat{\cal H}_{\rm 0} $ in the $U - V$ plane is sketched 
in Fig.~\ref{fig:diagram}. 
For small values of the nearest-neighbor repulsion there is a direct superfluid (SF) 
to Mott-insulator (MI) transition. 
For sufficiently large values of $U$, before entering a density-wave (DW) phase 
characterized by a non-uniform boson density with staggered ordering~\cite{DWorder},
the model defined by Eq.~(\ref{eq:EBH})
exhibits a second order quantum phase transition between two insulating phases~\cite{Tor06,Ber08,Ros12}, 
the distinction among them being protected by the bond-centered inversion symmetry 
of $\hat{\cal H}_0$~\cite{Gu09,Ber11,Pol12}.
Specifically, upon increasing $V$ the system evolves from a conventional MI 
to a bosonic Haldane insulator (HI), which displays similar correlations to the Haldane 
gapped phase~\cite{Hal83} of spin-1 chains (it is indeed identified by a non-vanishing string order 
parameter~\cite{Tor06}, defined according to ${\cal O}_{\rm string} = \lim_{r \to \infty} 
\langle \delta \hat{n}_j e^{i \pi \sum_{k=j}^{j+r} \delta \hat{n}_k} \delta \hat{n}_{j+r} \rangle$, 
where $\delta \hat{n}_j$ denotes the boson number fluctuations from the average filling). 
Notice however that while for the ring configuration, 
{\it i.e.}, for Periodic Boundary Conditions (PBC), the HI phase is associated to a single non degenerate 
ground state, for Open Boundary Conditions (OBC) it exhibits a four-fold degeneracy~\cite{Ken92} 
that needs to be accounted for, when designing the parameter loop (see~Sec.~\ref{sec:obc} for details). 
The MI-HI transition defines the region of interest for our analysis:
in the rest of the paper we will fix the on-site interaction to the value $U=5$, 
so that the associated critical point $\mathcal C$ occurs at $V_{\rm c} \approx 2.7$
(see filled circle and dotted line in Fig.~\ref{fig:diagram}).

\begin{figure}[t]
  \includegraphics[width=0.9\linewidth]{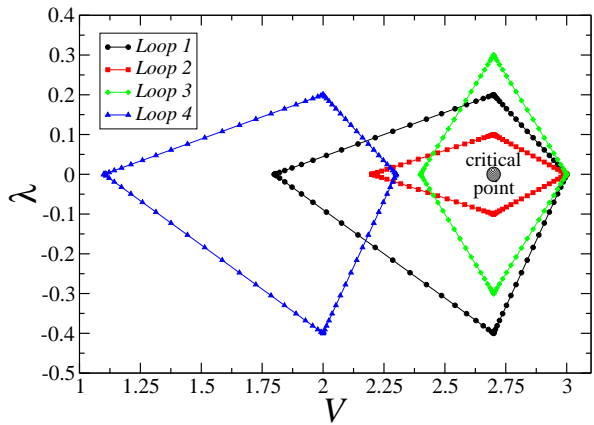}
  \caption{(color online). Parameter space of the Hamiltonian in Eq.~\eqref{eq:fullH} at $U=5$
    and unit filling. 
    The point $(V_{\rm c},0)$ with $V_{\rm c} =2.7$ is an isolated critical point separating two half lines, 
    $(V<V_{\rm c},0)$ and $(V>V_{\rm c},0)$, along which the system is in the MI and in the HI phase, 
    respectively [the same clear-cut distinction between the two phases being not possible 
      when $\lambda\neq 0$, owing to a lack of inversion symmetry of Eq.~(\ref{eq:lambda})].
    Different colors denote the loops $\Gamma$ that we have considered for the adiabatic 
    evolution of the system. 
    The loops marked with black circles ({\it ``Loop~1''}), red squares ({\it ``Loop~2''}), 
    and green diamonds ({\it ``Loop~3''}) enclose the critical point, and are thus expected 
    to be associated with a quantized number of pumped particles. 
    The loop with blue triangles ({\it ``Loop~4''}) is instead expected to give null pumped 
    charge, as it does not encircle~$(V_c,0)$.}
  \label{fig:loops} 
\end{figure}

As discussed in Ref.~\onlinecite{Ber08}, in order to have a non-zero pumped charge, one should 
drag the system adiabatically along a loop $\Gamma$ enclosing $\mathcal{C}$ in the parameter space 
of the Hamiltonian.
For this purpose, a second contribution $\hat{\cal H}_\lambda$ is added to Eq.~(\ref{eq:EBH}), 
which allows for an adiabatic connection between the MI and HI phases, by explicitly breaking 
the lattice inversion symmetry of~$\hat{\cal H}_0$. 
In particular, we choose a correlated hopping between neighboring sites which reads
\begin{equation}\label{eq:lambda}
  \hat{\cal H}_{\lambda} =  \lambda \sum_{j=1}^{L} (\hat{n}_{j} \, \hat{b}^{\dag}_{j} \hat{b}_{j+1} + {\rm h.c.})~,
\end{equation}
with $\lambda$ being a (controllable) intensity parameter.
With this choice, upon fixing $U$ as detailed above and by externally controlling $V$ and $\lambda$, 
we can now drag the system along a  pumping cycle  $\Gamma$ around the isolated critical point 
$(V,\lambda)=(V_{\rm c},0)$ that represents $\mathcal{C}$ in the $\lambda - V$  plane (see Fig.~\ref{fig:loops}).
Specifically, assuming the system to be initialized into the ground state $|\psi_{GS}\rangle$ 
of the Hamiltonian~\eqref{eq:fullH} evaluated for some given point $(V_0,\lambda_0)$ of $\Gamma$, 
we evolve it via a time-dependent Hamiltonian $\hat{\cal H}[V(t),\lambda(t)]$ obtained 
from~\eqref{eq:fullH} by varying parametrically in time the quantities $V$ and $\lambda$ 
along the loop $\Gamma$ (the trajectory being periodic with period $T$). 
Accordingly, at time $t$ the vector state becomes
\begin{equation} \label{NEWEQ111}
  |\psi(t)\rangle = \overleftarrow{\rm T} e^{-i\int_0^t \hat{\cal H}[V(t'),\lambda(t')] \, {\rm d}t'} 
  |\psi_{\rm GS}\rangle \,,
\end{equation}
with $(V(t),\lambda(t)) \in \Gamma$ for all $t$ and with $(V(0),\lambda(0)) = (V_0,\lambda_0)$. 
This is the state on which, as detailed in Sec.~\ref{sec:pump}, we are going to 
evaluate the pumped charge accumulated on a cycle.

\section{Methods}   \label{sec:method}

In order to study the many-body system in Eq.~\eqref{eq:fullH}, we employ a numerical approach based
on the t-DMRG method in the formalism of the matrix product states (MPS)~\cite{Sch11}. 
In particular, to characterize the properties of the ground state $\vert \psi_{\rm GS} \rangle$ 
of Eq.~(\ref{eq:fullH}) (such as the energy and the relevant correlation functions), 
we impose a MPS trial wave function, and perform a variational minimization of the energy 
cost function site by site.
Subsequently, in order to follow the evolution of the vector in Eq.~(\ref{NEWEQ111}),
we adopt the time-evolving block decimation (TEBD) procedure~\cite{vidal}.
In our simulations we were able to keep track of the globally conserved number of bosons,
thus always working in the canonical ensemble. We fixed the number of particles in order to have
unit filling ($\bar n=1$), and distributed them over lattices of up to $L=200$ sites~\cite{trunc}.

The choice of the boundary conditions of the chain requires a special attention: as a matter of fact, 
most of the time-dependent simulations presented in this paper (except those in Sec.~\ref{sec:obc}) 
have been carried out for a ring geometry, {\it i.e.}, under PBC.
It turns out that the standard t-DMRG algorithm is not optimized for these boundary configurations.
Indeed, on one hand this choice imposes a cyclic structure to the MPS ansatz~\cite{pbc_porras}, 
thus resulting in a considerable slowdown of the static optimization algorithm, 
unless some clever strategies are adopted~\cite{pbc}.
On the other hand, the inability to construct an orthonormal basis, starting 
from the leftmost and from the rightmost MPS sites and requiring the proper unitary conditions
on the defining matrices, prevents from the possibility to write any bipartition
of the system in two contiguous blocks of sites as a Schmidt decomposition.
This causes a failure in the standard TEBD scheme and leads to an uncontrollable truncation error,
which demands for a substantial increase of the required MPS bond-link dimension $m$
and a drastic slowdown of the computational performances.

\begin{figure}
  \includegraphics[width=1.0\linewidth]{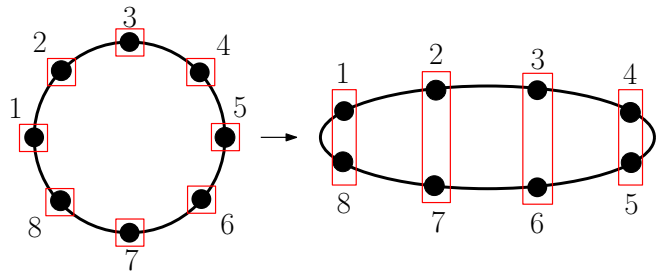} 
  \caption{(color online). Implementation of PBC using a OBC scheme for the underlying MPS ansatz.
    The left picture exemplifies the standard approach, in which each physical site (black circles)
    is associated to a computational site (red boxes), so that the MPS has a cyclic structure.
    The right picture denotes our scheme: to each physical site correspond two computational sites,
    thus effectively ``flattening'' the ring as an open-ended 1D shape.
    In the drawings we sketched a chain with $L=8$ sites.}
  \label{fig:dmrg_PBC} 
\end{figure}

Instead of directly imposing PBC, we found it more convenient to implement a standard t-DMRG scheme 
with OBC, where the 1D physical system with a ring geometry 
has been ``flattened'' as a narrow string and therefore mapped into an open ended chain.
As pictorially shown in Fig.~\ref{fig:dmrg_PBC}, we associate to each computational site 
two distant physical sites. 
This trick enables us to use a conventional t-DMRG code for open boundaries and to
take advantage of all the above mentioned methods for standard OBC.
In particular we are able to easily implement the conservation of abelian symmetries and 
to control the truncation error within the TEBD scheme.
The price we have to pay is that of doubling the local Hilbert space of each computational site, and 
using a larger $m$ value in order to get accuracies comparable to the OBC situation~\cite{note_bondlink}.

Our code dynamically updates the bond-link dimension used at each link, 
both in the static algorithm and during a single step of time evolution,
according to a given threshold value in the sum of the discarded weights $\varepsilon$. 
We found that, in order to keep it below $5 \times 10^{-8}$, a number of states $m \lesssim 400$ 
is generally sufficient~\cite{note_adiab}.
We checked that a further reduction of $\varepsilon$ does not significantly affect our results.
In the TEBD dynamics, we employed a fourth-order Trotter expansion of the time-evolution operator,
thus minimizing the error due to a finite time step (but still admitting time steps of the order
$dt \sim 5 \times 10^{-2}$).

\section{Energy gaps}   \label{sec:gaps}

As we are dealing with finite-size systems characterized by the Hamiltonian in Eq.~\eqref{eq:fullH},
the behavior of the ground-state energy gap is important, especially at the critical point 
and along the loop in parameter space. 
In order to enter the adiabatic regime, ideally one would like to complete the loop 
in a time $T$ which is much larger than the inverse of
\begin{equation}
  \Delta_{sm} = \min_{(V,\lambda)\in \Gamma} \Delta_{V,\lambda}\;,
\end{equation}
that is the smallest of the gaps $\Delta_{V,\lambda}$ of the Hamiltonians $\hat{\cal H}[V,\lambda]$ 
evaluated along the selected loop~$\Gamma$.
In addition, for a finite-size system, the gap $\Delta_c$ of the Hamiltonian~(\ref{eq:fullH}) 
at the critical point $(U=5, V_c\simeq 2.7, \lambda=0)$ is never exactly zero. 
Therefore, on decreasing the amplitude of the loop enclosing the singularity, one should start 
to observe deviations from the quantized value for the smallest loops for which the finiteness 
of the critical gap starts to be resolved. 
An additional constraint on the amplitude of the pumping path raises due to the emergence of the DW phase 
close to the Haldane insulator.  

All these requirements lead us to consider the loops shown in Fig.~\ref{fig:loops}, 
where we depict four paths in the $\lambda - V$ parameter space 
along which we have computed the evolution governed by the time-dependent 
Hamiltonian $\hat{\cal H}(t) \equiv \hat{\cal H}[V(t),\lambda(t)]$ with period 
$T$: $\hat{\cal H}(t+T) = \hat{\cal H}(t)$. Three of them (black, red, and green set) 
encircle the critical point $(V_{\rm c},0)$, while the last one (blue set) is trivial, 
meaning that the gap is never vanishing inside it.
The asymmetric shape of the loops with respect to the critical point is a trade off 
between the need to avoid the influence of the nearby DW phase, and at the same time 
to have a contour sufficiently far from the critical point in order to minimize 
the effects of the finiteness of the critical gap $\Delta_c$.

In order to define the time scales on a more quantitative basis, in the following 
we are going to analyze the scaling of the gaps $\Delta_c$ and $\Delta_{V,\lambda}$ 
as a function of the system size $L$. 
To do so, we focus on the ground-state charge gap, that is the difference in the energy needed 
to add ($\Delta E^+$) or to remove ($\Delta E^-$) a single boson 
in the system: $\Delta_{\rm charge} = \Delta E^+ - \Delta E^-$.
In the whole MI, as well as in the HI phase far from the transition with the DW phase~\cite{Tor06},
the lowest energy excitations of the Hamiltonian~(\ref{eq:fullH}) are indeed charge excitations, 
therefore the values of $\Delta_{\rm charge}$ (evaluated for the proper $V$ and $\lambda$ values) 
can be used to derive $\Delta_c$ and $\Delta_{V,\lambda}$.

\begin{figure}
  \includegraphics[width=0.9\linewidth]{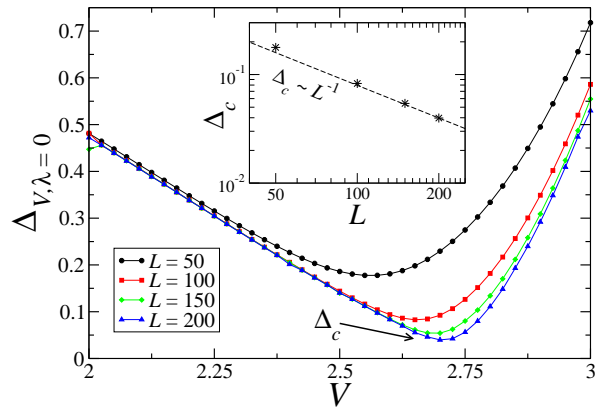}
  \caption{(color online). Finite-size scaling of the charge gap $\Delta_{V,\lambda}$ 
    at $\lambda = 0$, close to the critical point and for $U=5$. 
    At the phase transition, $V_c \approx 2.7$, the gap scales to zero as $L^{-1}$
    (see the inset, where we plot the minimum reached value $\Delta_c$ as a function of 
    $L$---the dashed line indicates a behavior $\Delta_c \sim 1/L$ and is plotted as a guide to the eye). 
    Away from criticality, the gap is almost independent on the length, as long as $L \gtrsim 100$. 
    As detailed in text, the values of $\Delta_{V,\lambda=0}$ and $\Delta_c$ reported here 
    have been computed by evaluating the charge gap $\Delta_{\rm charge}$ of the system.}
  \label{fig:critgap} 
\end{figure}

Let us first discuss the gap close to the critical point, $\Delta_c$.
In Fig.~\ref{fig:critgap} we plot the size dependence of the charge gap, 
and show that, in the thermodynamic limit, it vanishes as $1/L$.
For our convenience we computed the gaps in a chain with OBC. The differences are not relevant for 
our purposes ({\it i.e.}, to determine the relevant time-scales for the pumping period). On the other 
hand in this way we could easily analyze the critical gap for chains up to few hundred sites~\cite{Ros12}.

\begin{figure}
  \includegraphics[width=0.9\linewidth]{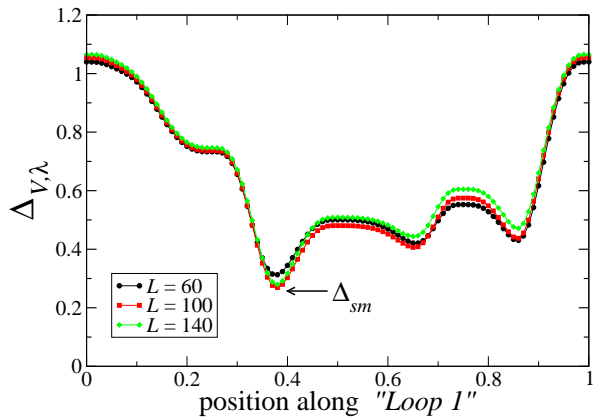}
  \caption{(color online). Finite-size scaling of the gap $\Delta_{V,\lambda}$ along {\it ``Loop~1''} 
    (see Fig.~\ref{fig:loops}) for three different system sizes, $L=60, 100, 140$ 
    [as in Fig.~\ref{fig:critgap}, the values of $\Delta_{V,\lambda}$ have been obtained 
      by computing the charge gap $\Delta_{\text{charge}}$].
    The minimum value of the gap $\Delta_{sm}$ (indicated by an arrow in the figure) 
    is the bottleneck to define the adiabatic dynamics.}
  \label{fig:loopgap} 
\end{figure}

Now the interesting point which permits us to address the regime of topological pumping
is that, already for $L \sim 200$, the gap $\Delta_c$ turns out to be almost one order 
of magnitude smaller than the typical gap $\Delta_{V,\lambda}$ encountered along the loops $\Gamma$ 
of Fig.~\ref{fig:loops}.
A quantitative analysis of the value of the gap along one particular loop 
is shown in Fig.~\ref{fig:loopgap}, where we took data directly in the case with PBC 
(specifically, here we analyzed {\it ``Loop~1''}). 
In accordance with the fact that along the loop the system remains distant from criticality, 
the gap barely depends on the size.
We notice that the value of $\Delta_{V,\lambda}$ is not uniform along the loop 
and can significantly diminish at some points. 
This imposes strict conditions on the evolution period $T$ necessary to observe quantized pumping 
and calls for a careful analysis of the role of non-adiabatic corrections. 

\subsection{Testing the adiabatic condition} 

Roughly speaking, one could expect that the adiabatic regime is achieved for periods 
$T \gg \Delta_{sm}^{-2}$.
A good estimator of the quality of the adiabatic approximation along the performed loop
is given by the excess energy $\Delta E_{\rm exc}(t)$. This is quantified 
by the difference between the energy of the wavefunction during the evolution, 
$\langle \psi(t) \vert \hat{\cal H}(t) \vert \psi(t) \rangle$, 
and the energy $\langle \psi_{\rm GS}(t) \vert \hat{\cal H}(t) \vert \psi_{\rm GS}(t) \rangle$ 
corresponding to the instantaneous ground state $\vert \psi_{\rm GS}(t) \rangle$
of $\hat{\cal H}(t)$, {\it i.e.},
\begin{equation}
  \Delta E_{\rm exc}(t) = \langle \psi(t) \vert \hat{\cal H}(t) \vert \psi(t) \rangle 
  - \langle \psi_{\rm GS}(t) \vert \hat{\cal H}(t) \vert \psi_{\rm GS}(t) \rangle \;\; . 
\end{equation}

\begin{figure}
  \includegraphics[width=0.9\linewidth]{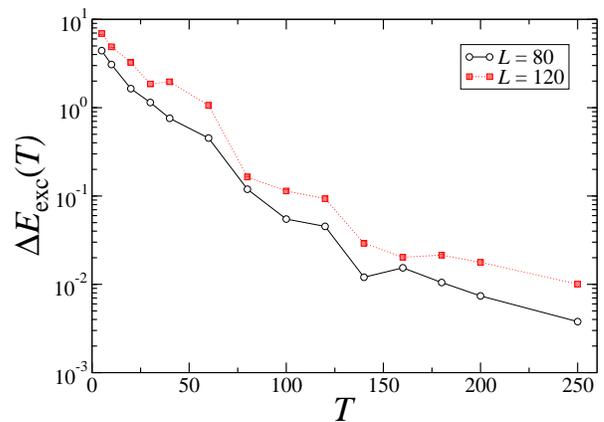}
  \caption{(color online). Excess energy, with respect to the actual adiabatic 
    ground state, evaluated at the end of the loop, {\it i.e.}, once the system parameters 
    have been restored to the initial values. Here $T$ denotes the total evolution time.
    The data have been obtained by performing a cyclic variation of the parameters 
    $(V, \lambda)$ along {\it ``Loop~1''}, starting from the upper point 
    in the $\lambda-V$ plane.}
  \label{fig:Energy_T} 
\end{figure}

In Fig.~\ref{fig:Energy_T} we analyzed the value $\Delta E_{\rm exc}(T)$ at the end 
of the loop, as a function of the total evolution time $T$.
The excess energy is generally decreasing with $T$, thus meaning that the adiabaticity
condition along the loop can be progressively approached by slowing down the variation 
of the system parameters $(\lambda, V)$.
Moreover, we note that $\Delta E_{\rm exc}(T)$ is not strongly affected by
the system size $L$. This is consistent with the fact that the instantaneous gap 
barely depends on $L$, provided this is sufficiently large (see Fig.~\ref{fig:loopgap}).
We finally point out that, far from the adiabaticity condition, such energy is not
guaranteed to be monotonic in $T$.

\section{Pumped charge}   \label{sec:pump}

Let us now introduce the quantity of interest, the pumped charge, to be evaluated by means 
of direct TEBD numerical approach for the time evolution of 1D many-body systems.

The number of particles which have flown through site $j$ between time $t$ and $t'>t$ 
can be expressed as the integral of the instantaneous 
local bosonic current ${\cal J}_{j}(t)$ at site $j$:
\begin{equation}
  {Q}^{(j)}_{t,t'} = \int_{t}^{t'} {\cal J}_{j}(\tau)~d\tau~,
  \label{eq:chargeInst}
\end{equation}
with ${\cal J}_{j}(t) = \langle \psi(t) \vert \hat{\cal J}_j(t) \vert \psi(t) \rangle$ 
being evaluated as the expectation value of the local current operator $\hat{\cal J}_j(t)$. 
Notice that, even though Eq.~(\ref{eq:chargeInst}) is expressed in the Schr\"odinger picture, 
the operator $\hat{\cal J}_{j}(t)$ depends on $t$ via the parametric temporal dependence 
of $V$ and $\lambda$ entering the system Hamiltonian. 
Explicitly, such operator is determined via the continuity equation
\begin{equation}
  \hat{\cal J}_j(t) - \hat{\cal J}_{j-1}(t) = i \big[ \hat{n}_{j},\hat{\cal H}(t) \big]~,
\end{equation}
yielding an expression, in terms of two-point correlators on nearest neighbor sites, of the form
\begin{equation}
  \hat{\cal J}_j(t) = i \big( - \hat{b}^\dagger_j \hat{b}_{j+1} 
                                 + \lambda \, \hat{n}_j \hat{b}^\dagger_j \hat{b}_{j+1} - {\rm h.c.} \big)~.
\end{equation} 

When integrating over a period $T$, the quantity~(\ref{eq:chargeInst}) gives the number 
of pumped bosons per cycle. Under the conditions of topological pumping, this quantity 
is independent of the initial time $t$ and of the starting point inside the loop, {\it i.e.}, 
\begin{equation}
  {Q}^{(j)}_{t,t+T} = {Q}^{(j)}_{0,T} \equiv Q_j(T)  = \oint_{\rm loop} {\cal J}_{j}(\tau)~d\tau \;,
  \label{eq:chargeT}
\end{equation}
furthermore, owing to the equivalence between lattice sites, for a system with PBC,
$Q_j(T)$ does not depend on $j$.

In evaluating the function~(\ref{eq:chargeT}), we are going to consider two different configurations: 
a ring with periodic boundary conditions, and a finite chain with open boundary conditions.
In the first one (Sec.~\ref{sec:ring}), the periodic modulation 
of the external parameters $(V(t), \lambda(t))$ induces a uniform 
and persistent current ${\cal J}(t)$ that is independent of the position $j$. 
The pumped charge $Q$ is the current passing through a link integrated over a period $T$. 
In the second setup (Sec.~\ref{sec:obc}), the boundaries break the translational invariance
and the instantaneous current ${\cal J}_j (t)$ is actually a site-dependent quantity.
As we will show below, it turns out that the pumping mechanism tends to deplete 
the particle density on one side, and to increase it on the opposite one. 
This progressively destroys the insulator, which can be stabilized only 
at a commensurate filling, starting from the edges.
We measure the current, and hence the pumped charge, close to the middle part of the chain,
where the filling remains close to unity for the whole duration of the cycle.

\section{Ring configuration}   \label{sec:ring}

As already mentioned in the previous section, for a system with PBC, the pumped charge 
on a cycle, Eq.~(\ref{eq:chargeT}), does not depend on $j$, {\it i.e.}, $Q_j(T) = Q(T)$. 
As a matter of fact, the same property applies for all the quantities defined 
in Eq.~(\ref{eq:chargeInst}), yielding ${Q}^{(j)}_{t,t'} = {Q}_{t,t'}$.

\subsection{Finite-size scaling of the pumped charge}   \label{finitesize}

\begin{figure}[t]
  \includegraphics[width=0.9\linewidth]{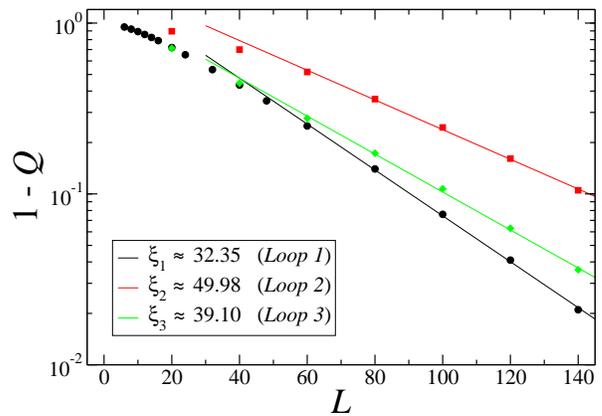} 
  \caption{(color online). Ring configuration: number of pumped particles $Q$ per cycle as a function 
    of system size $L$, corresponding to the three different loops encircling the critical point
    (black circles stand for {\it ``Loop~1''}, red squares for {\it ``Loop~2''}, 
    while green diamonds for {\it ``Loop~3''} -- see Fig.~\ref{fig:loops}). 
    In our simulations we used different periods $T$ according to the size,
    such to ensure the adiabatic approximation along the whole loop
    ({\it e.g.}, we took $T=100$ for $L=8$, and $T=350$ for $L=140$).
    Data are displayed in a logarithmic plot. The straight lines are best fits 
    to the numerical data for $L \geq 60$, and are respectively given by 
    $1-Q \approx 1.64 \times e^{-0.031 \, L}$ ({\it ``Loop~1''}),
    $1-Q \approx 1.76 \times e^{-0.02  \, L}$ ({\it ``Loop~2''}), 
    $1-Q \approx 1.32 \times e^{-0.026 \, L}$ ({\it ``Loop~3''}).}
\label{fig:charge_L} 
\end{figure}

Our numerical results for $Q(T)$ after one cycle with a sufficiently long period $T$, 
in the sense discussed in the following Subsec.~\ref{nonadiabatic},
and enclosing an isolated quantum critical point are summarized in Fig.~\ref{fig:charge_L}.
There we report data (depicted by black circles, red squares and green diamonds) 
corresponding to the three different loops around the critical point in Fig.~\ref{fig:loops}, 
as a function of the system size $L$~\cite{thermlim}. 
We note that the total pumped charge approaches the unit value for sufficiently large 
system sizes, no matter how we choose the specific path encircling the critical point. 
This is an important observation supporting the topological origin of the quantization of $Q$, 
which, strictly speaking, is valid only in the thermodynamic limit. 
Remarkably, it rules out the possibility that $Q$ remains simply a geometric quantity, 
even in the limit of an infinite chain, while instead it is has to be taken as a topological 
and quantized property of the system.
As long as the size $L$ of the system is finite, we have discrepancies 
in the final value $Q$ at the end of the cycle computed for different paths. 
Such discrepancy decreases with increasing number of lattice sites and vanishes 
in the thermodynamic limit, when $Q$ is quantized to unity, independently 
of the specific path considered.

A deeper investigation of the scaling properties of $Q$ with the system size 
reveals corrections to the topological value that decrease exponentially, as witnessed by the quality of the fits 
to the numerical data (symbols) that are shown in Fig.~\ref{fig:charge_L}.
Our fits indicate that, for all the three loops and for sufficiently large sizes, 
\begin{equation}
  Q \approx 1 - e^{-L/\xi}~,
\end{equation}
where the decay rate $\xi$ depends on the specific path.
In particular we found $\xi_1 \approx 32.35$ for {\it ``Loop~1''},
$\xi_2 \approx 49.98$ for {\it ``Loop~2''},
and $\xi_3 \approx 39.10$ for {\it ``Loop~3''}.
Although a detailed analysis of the dependence of $\xi$ on the chosen loop
is computationally demanding and lies beyond the purposes of the present paper,
we observe that $\xi$ increases when the loop shrinks around the critical point.
This is apparent by looking at Fig.~\ref{fig:loops} and noticing that,
while {\it ``Loop~2''} is entirely contained in {\it ``Loop~1''},
we found $\xi_2 > \xi_1$. In addition, this is in agreement with the fact that 
the larger the loop, the smaller the corrections to the quantized value of the pumped charge 
due to the finiteness of the critical gap $\Delta_{c}$ when $L$ is finite. 

\begin{figure}
  \includegraphics[width=0.9\linewidth]{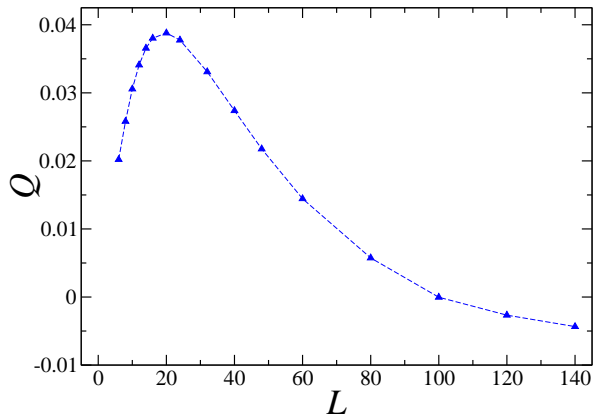} 
  \caption{(color online). Same as in Fig.~\ref{fig:charge_L}, but for {\it ``Loop~4''} (according 
    to the notation of Fig.~\ref{fig:loops}), which does not enclose the critical point.}
  \label{fig:charge_nocrit} 
\end{figure}

On the other hand, when the loop does not enclose a critical point, the number of pumped particles 
per cycle goes to zero in the limit $L\to \infty$, as shown in Fig.~\ref{fig:charge_nocrit}
for {\it ``Loop~4''}.
The actual asymptotic value at the thermodynamic limit is not exactly zero, as it is
apparent from the data. However we point out that this discrepancy may be due to numerical 
inaccuracies (an error in $Q$ of the order of $\varepsilon \lesssim 0.005$ 
is negligible on the logarithmic scale used for the data points in Fig.~\ref{fig:charge_L}).

The achievement of perfect quantization ($Q=0$ or $Q=1$) only in the 
thermodynamic limit is consistent with the topological origin of the pumping mechanism. 
As long as $L$ remains finite, the gap at the critical point $\Delta_{c}$ is never 
exactly zero and the topological argument for quantization does not hold. 
We have been able to observe quantization as long as the loop encircles the critical point, provided 
it lies at a sufficient distance from it. We do expect that, on choosing loops of decreasing amplitude, 
we should start observing deviations from the quantized value even for large (but still finite) $L$. 
For small loops, the number of pumped bosons per cycle, still retaining its geometrical origin, 
looses to be quantized unless we go to very large system sizes. 
Although it would have been interesting to show this cross-over, there are numerical limitations 
which prevent us to consider loops with amplitudes smaller than the {\it ``Loop 2''} 
of Fig.~\ref{fig:loops}. 
Reducing the amplitude of the loop implies a longer time $T$ needed to achieve the adiabatic 
regime, much larger than $T \sim 300$ used to study the cycles of Fig.~\ref{fig:loops}. 
This is outside our numerical capabilities.

\subsection{Non-adiabatic corrections}   \label{nonadiabatic}

Let us now examine in more details the non-adiabatic effects on the pumped charge.

\begin{figure}
  \includegraphics[width=0.95\linewidth]{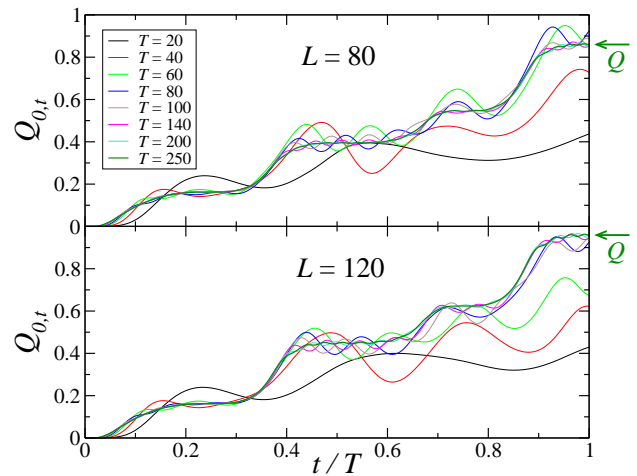}
  \caption{(color online). Time-resolved number of pumped particles $Q_{0,t}$ 
    as a function of time, in units of the evolution period $T$. 
    Different curves report results for different values of $T$, 
    while the upper (lower) panel shows data for a system of $L=80$ ($L=120$) sites.
    The data have been obtained choosing {\it ``Loop~1''} and starting from the 
    upper point in the $\lambda-V$ plane.
    The two right arrows indicate the final pumped charge $Q$ as extracted from the 
    large-$T$ set of data (and eventually shown in Fig.~\ref{fig:charge_L}).}
\label{fig:charge_T} 
\end{figure}

The time-resolved number of pumped particles $Q_{0,t}$
for a given finite pumping period $T$ typically exhibits wiggles in time,
which disappear with increasing $T$.
In order to ensure the validity of the adiabatic approximation along the loop, 
one has to take a total evolution time much larger than the inverse square 
of $\Delta_{sm}$. 
Once this condition is satisfied, the total number of transported bosons $Q=Q(T)$ on a cycle
is almost independent of $T$, thus promoting $Q$ to a geometric quantity. 
This is apparent from Fig.~\ref{fig:charge_T}, where we show $Q_{0,t}$ 
as a function of time $t$ (in units of $T$), for different values 
of the evolution period $T$ and for two different sizes. 
The asymptotic value of the total pumped charge is approached for large $T$,
while the non-adiabatic wiggles tend to disappear.

\begin{figure}
  \includegraphics[width=0.9\linewidth]{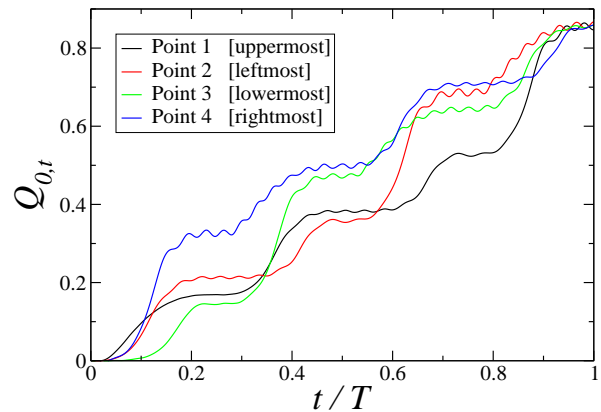}
  \caption{(color online). Time-resolved number of pumped particles $Q_{0,t}$ as a function 
    of time (in units of $T$). We show results for different initial points 
    along the path marked as {\it ``Loop~1''}, corresponding to the four extremal points
    on the $V$- and the $\lambda$-axes. 
    These data are for $L=80$ sites and for an evolution time $T=200$.}
  \label{fig:charge_pos} 
\end{figure}

  \begin{figure*}[!t]
    \includegraphics[width=0.8\linewidth]{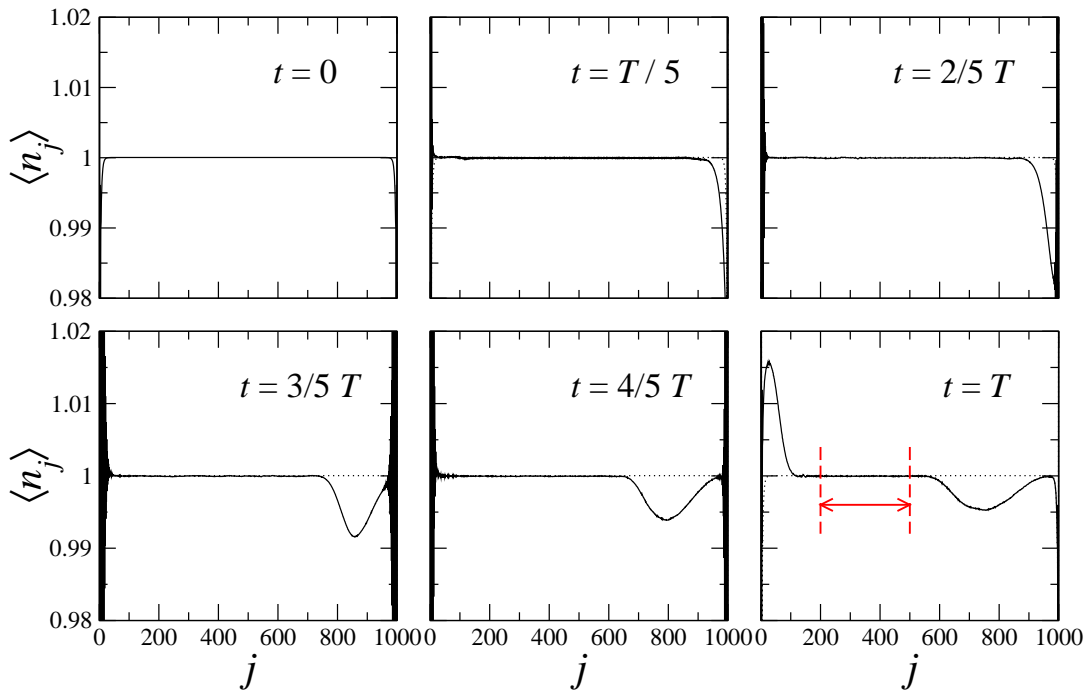}
    \caption{(color online). Average number of particles at each site of an open-ended chain, 
      described by the Hamiltonian in Eq.~\eqref{eq:fullH}.
      Different panels stand for different times during an adiabatic variation of
      the parameters along the path marked as {\it ``Loop~1''}.
      Here we chose, as initial ground state, the leftmost point of the loop 
      corresponding to a pure MI state (see Fig.~\ref{fig:loops}), and then 
      follow the path in the anti-clockwise direction.
      The data refer to a $L=1000$ chain, and to an evolution time $T=300$.
      The red arrow in last panel denotes the surviving insulating region 
      after one adiabatic loop.}
    \label{fig:density_obc} 
  \end{figure*}

Finally we note that, once the adiabatic condition along the loop is satisfied, 
the total pumped charge does not depend on the initial point along the loop,
as explicitly shown in Fig.~\ref{fig:charge_pos}.

We conclude this section by observing that the plateaus emerging
in the time-resolved pumped charge are adiabatic, but of a non topological nature, 
and depend on the fine details of the instantaneous low-lying energy spectrum.
They can change according to the specific shape 
of the loop around the critical point (Fig.~\ref{fig:charge_T}), 
as well as on the initial condition along the loop (Fig.~\ref{fig:charge_pos}).

\section{Pumping in an open-ended chain}   \label{sec:obc}

Quantum pumping can also be studied in an open-ended chain.
In this case, differently from the ring geometry, the system no longer exhibits translational 
invariance and all the quantities defined in Sec~\ref{sec:pump} become site-dependent.
In Sec.~\ref{sec:obc_results} we display numerical evidence of the fact that this configuration leads to 
a pumping mechanism which, after one loop, produces the transport of a quantized 
bosonic charge from one end to the opposite end of the chain.

We remark that, in order to guarantee unit filling everywhere in the chain
[modelled by Eq.~(\ref{eq:fullH})] during the whole cycle, it would be necessary 
to couple the system to two (non-interacting) reservoirs attached at the two ends.
In the absence of any direct bias between the electrodes, the charge would be transferred by 
a uniform modulation of the parameters $\lambda$ and $V$ in the middle (interacting) part 
of the system, along the loops like those depicted in Fig.~\ref{fig:loops}.
On the other hand, in order to prevent a depletion of particles in the interacting region
in favor of the non-interacting leads, it would be necessary to consider a regime
of weak system-reservoir coupling, thus finding a trade-off between depletion and relaxation 
of pumped particles into the leads. For interacting bosonic systems, this regime is actually 
very difficult to achieve, not only in simulations, but even conceptually.
For this reason, and owing to the fact that experimentally it would be much easier to realize 
an open-ended chain, we have focused on a system with OBC.

\subsection{Numerical results}   \label{sec:obc_results}

As shown in Fig.~\ref{fig:density_obc}, one immediately recognizes that, 
during the time evolution ruled by the adiabatic variation of the parameters $(\lambda,V)$ 
along a closed loop (see Fig.~\ref{fig:loops}), the average on-site density changes with time. 
In particular, following {\it ``Loop~1''} in the anti-clockwise direction, we are able 
to progressively detect a net imbalance of bosonic particles from the right end, 
where we observe a depletion from the average unit filling, to the left end, 
where a bump in the density profile eventually appears.
In order to probe the quantization of particle transport, one then has to investigate 
the charge ${\cal Q}^{(j)}_{0,t}$ pumped along one cycle, according to Eq.~\eqref{eq:chargeInst}. 

Before discussing our results, it is important to stress that, for OBC, the charge transport
from one edge to the opposite edge of the chain unavoidably drives the system
out of equilibrium, so that, for any $t>0$, the many-body wavefunction $\vert \psi(t) \rangle$ 
cannot be in the instantaneous ground state (ultimately this is due to the fact that, 
when passing through the HI phase, the system faces a four-fold degenerate
ground-state). This contrasts with the translationally invariant PBC setup,
where the adiabatic approximation can be satisfied for sufficiently slow time variations
of the Hamiltonian parameters (see Sec.~\ref{sec:gaps}), and is associated to
the geometric nature of the pumping mechanism.
Interestingly enough however, despite adiabaticity cannot be fulfilled as a global condition, 
it can be still satisfied ``locally''. 
More precisely, this means that, even though the global state of the bosons might depart from the 
corresponding instantaneous ground-state of the driving Hamiltonian, the reduced density matrix 
associated with a specific collection of sites remains sufficiently close to its adiabatic counterpart.
Clearly, any deviation of the density profiles from unit filling indicates the loss of local adiabaticity.
Therefore, in order to satisfy such condition, the instantaneous current ${\cal J}_j(t)$ 
has to be computed in the bulk of the system, where the particle density is still
commensurate and the depletion/accumulation mechanism has not spoiled the insulating properties 
of the bosonic system. 
This requirement forces one to consider quite large sizes, since the region
of deviation from the unit filling is proportional to the period $T$ and involves a 
non negligible part of the whole system, in all the simulatable cases.
We observed that a length of $L = 1000$ sites is adequate to our purposes~\cite{note_obc}.

  \begin{figure}[t]
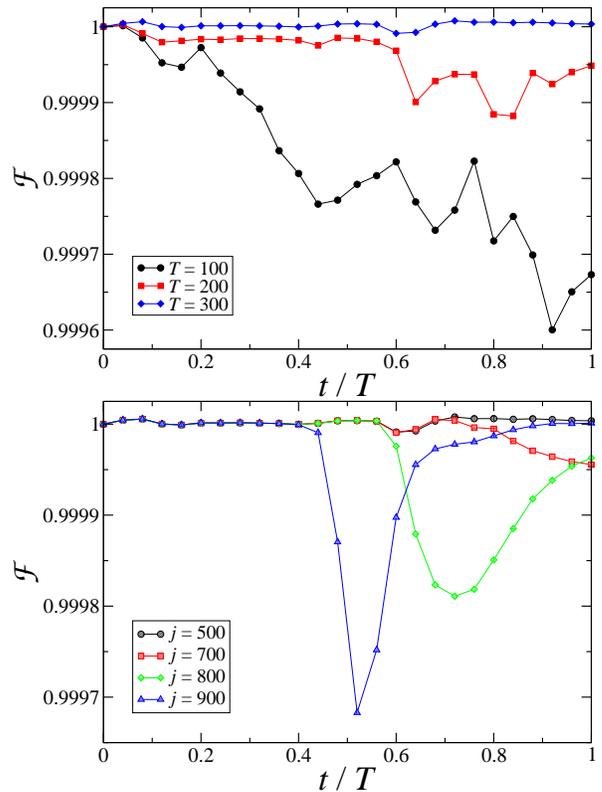

    \includegraphics[width=0.9\linewidth]{Fidelity_T}
    \includegraphics[width=0.9\linewidth]{Fidelity_j}
    \caption{(color online). Two-site nearest-neighbor fidelity between the instantaneous state 
      during the time evolution and the corresponding adiabatic ground state for a chain with $L=1000$. 
      (Top panel) The fidelity for two neighbouring sites in the centre of the chain is analysed 
      as a function of the period $T$. For $T=300$ the deviation from adiabaticity 
      is below $10^{-5}$ during the whole evolution.
      (Bottom panel) The fidelity is analysed as a function of the position of the link in the chain 
      for the largest period ($T=300$).
      The observed deviations close to the end of the chains are in correspondence with the deviations 
      from unit filling visible in Fig.~\ref{fig:density_obc}.}
    \label{fig:Fidelity_obc} 
  \end{figure}

We checked the validity of the adiabatic regime locally, by evaluating the fidelity ${\cal F}$ 
between the reduced density matrix of two nearest neighbor sites during the time evolution, 
$\hat{\rho}_{j,j+1}(t) = {\rm Tr}_{L - \{ j,j+1 \}} \big( \vert \psi(t) \rangle \langle \psi(t) \vert \big)$,
and the corresponding one on the instantaneous ground state,
$\hat{\rho}_{j,j+1}'(t) = {\rm Tr}_{L - \{ j,j+1 \}} \big( \vert \psi_{\rm GS}(t) \rangle \langle \psi_{\rm GS}(t) \vert \big)$.
Omitting the subscripts ${}_{j,j+1}$, this is given by 
\begin{equation}
  {\cal F}(t)= {\rm Tr} \left[ \sqrt{ \sqrt{\hat{\rho}'(t)} \, 
      \hat{\rho}(t) \, \sqrt{\hat{\rho}'(t)} } \, \right] \,.
  \label{eq:twosite}
\end{equation}
The fidelity ${\cal F}$ is a real number which can assume any value in the interval $[0,1]$,
and measures the distance between the two quantum states. If ${\cal F} \approx 1$,
the two states are nearly the same, up to some global phase factor. On the other hand,
if ${\cal F} \approx 0$, the two states are nearly orthogonal~\cite{NielsenChuang}.
In Fig.~\ref{fig:Fidelity_obc} we plotted ${\cal F}(t)$ for two nearest neighbor sites 
in the central part of the chain for different values of the period $T$ (top panel), and for the largest 
period as a function of the position $j$ of the link in the chain (bottom panel). 
For a total evolution time $T=300$, ${\cal F}(t) \approx 1$ along the whole loop, thus providing 
evidence of adiabaticity, in the central region (top panel of Fig.~\ref{fig:Fidelity_obc}). 
By departing from the middle, even at $T=300$ adiabaticity is lost (see the bottom panel 
of Fig.~\ref{fig:Fidelity_obc}). The results for the pumped charge reported below are taken 
in the region close to $j=500$, where adiabaticity is ensured during the whole cycle.

  \begin{figure}[!t]
    \includegraphics[width=0.9\linewidth]{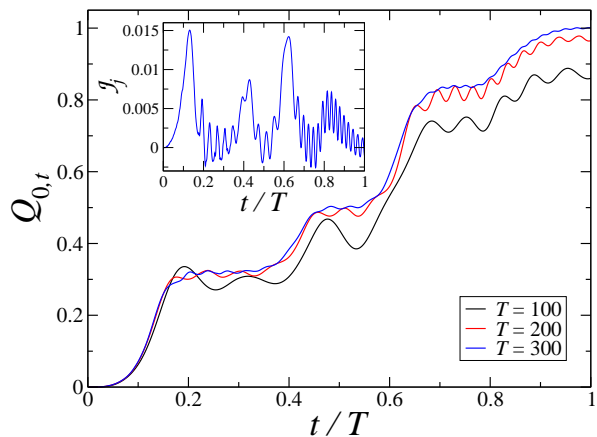}
    \caption{(color online). Time-resolved number of bosons which are pumped from the right 
      to the left side of the open-ended chain, as obtained from Eq.~\eqref{eq:chargeInst}.
      The various continuous lines refer to a system of $L=1000$ sites,
      for three different evolution periods $T$ along {\it ``Loop~1''},
      starting from the pure MI point and going in the anti-clockwise direction.
      The inset displays the instantaneous current ${\cal J}_{j}(t)$ averaged
      over the insulating region highlighted in Fig.~\ref{fig:density_obc}, for $T=300$.}
    \label{fig:charge_obc} 
  \end{figure}

Following this prescription, we averaged the current over the sites marked by 
the red arrow in the last panel of Fig.~\ref{fig:density_obc}, where we saw deviations 
from $\bar{n} = 1$ of up to $10^{-4}$, and the local adiabatic condition is satisfied.
The results are shown in Fig.~\ref{fig:charge_obc}, for three different values of
the evolution period $T$ along {\it ``Loop~1''} in the anti-clockwise direction.
For the maximum $T$ we considered ($T=300$), we observe a remarkably high level of accuracy 
in the quantization of the pumped charge. For lower values of $T$, perfect
quantization is lost and wiggles signaling adiabaticity losses start to emerge,
in analogy with the ring geometry already discussed in Sec.~\ref{nonadiabatic}.
To check the consistence of this result, we also computed the total number
of particles on the left and on the right of the insulating region, finding that
deviations from half filling correspond, at the end of the cycle, to having
respectively one particle more and one particle less.

Finally we simulated the pumping protocol by inverting the travelling direction 
along the loop, and found a scenario qualitatively similar to that of Fig.~\ref{fig:density_obc},
with the depleted and the filled edges inverted.
The corresponding pumped charge behaves analogously, despite the pumping mechanism 
is working in the opposite direction. We point out that the two directions are not symmetric, 
while only the total number of pumped bosons is quantized in the two cases.

\section{Conclusions}   \label{conclusions}

Adiabatic pumping in interacting systems is still a rather unexplored field, 
especially in the case of bosonic particles. 
When the competition between different order parameters gives rise to a quantum phase transition, 
the adiabatic circumnavigation of the critical point in  parameter space leads 
to a quantized number of pumped particles with topological protection. 

A possible realization has been proposed by Berg {\it et al.}~\cite{Ber11} 
and consists of an extended Bose-Hubbard model in one dimension in the presence 
of an inversion-symmetry-breaking term. Here we have investigated this model by means 
of time-dependent density matrix renormalization group simulations, giving numerical support 
to their original prediction. We have computed the pumping properties of the system 
as a function of the system size and shown that the scaling is consistent with a number 
of pumped bosons per cycle equal to one or zero, depending on whether the loop performed 
in parameter space encircles or not the critical point. 
In addition, we have provided evidence that the pumping mechanism is indeed adiabatic 
and that it displays topological protection by showing that, in the thermodynamic limit, 
the number of pumped particles is independent of both the total evolution time $T$ 
and of the specific path performed in parameter space. 
We considered two different configurations: a ring geometry, such that a persistent current 
arises as a consequence of the pumping mechanism, and an open chain, where quantum pumping 
is manifested in a depletion of charge at one end of the chain with a corresponding accumulation 
at the opposite end.

\acknowledgments
We are grateful to E. Altman, S. Peotta and F. Taddei for their useful comments. 
The work was supported in part by EU FP7 Programme under Grant Agreement No. 238345-GEOMDISS, 
No. 234970-NANOCTM, and No. 248629-SOLID.


\begin{thebibliography}{77}
\bibitem{DWorder}
  The DW phase is revealed by the asymptotic behavior of the correlator 
  ${\cal O}_{\rm DW} = \lim_{r \to \infty} (-1)^r \langle \delta \hat{n}_j \delta \hat{n}_{j+r} \rangle$, 
  where $\delta \hat{n}_j$ denotes the boson number fluctuations from the average filling.
\bibitem{note_adiab}
  Addressing the time evolution at long times is generally a hard task for any numerical approach.
  With t-DMRG methods, the required number of states tends to increase exponentially with time.
  This problem is especially relevant when the higher part of the Hamiltonian 
  excitation spectrum becomes relevant.
  In our case, the fact that we are simulating a system which is always close 
  to its instantaneous ground state and is well protected by a gap (adiabatic condition), 
  drastically helps in keeping $m$ down to a relatively small value.
\bibitem{thermlim}
  We notice that, since we consider $\bar n=1$, $L$ also coincides with the number of bosons. 
  As a consequence the system becomes truly a many-body system in the thermodynamic limit.
\bibitem{note_obc}
  The open-ended geometry enables us to use the standard DMRG approach for OBC,
  which in general can be simulated much more efficiency than PBC.
  We are able to safely simulate the adiabatic dynamics of a gapped system 
  with up to $L \sim 10^3$ sites, by using small bond-link dimensions, and yet
  achieving an accuracy in the total number of discarded weight of $\varepsilon \lesssim 10^{-8}$
  (in the specific, Figs.~\ref{fig:density_obc}-\ref{fig:charge_obc} 
  refer to $L=1000$ sites using $m \leq 120$ and a fourth-order trotter expansion
  of the evolution operator with $dt = 4 \times 10^{-2}$).
\end{thebibliography}
\end{document}